\documentclass[12pt]{article}
\usepackage{graphicx}
\usepackage{amsmath}
\usepackage{amssymb}

\def\beq{\begin{equation}}
\def\eeq#1{\label{#1}\end{equation}}
\def\eeqn{\end{equation}}

\def\beqa{\begin{eqnarray}}
\def\eeqa#1{\label{#1}\end{eqnarray}}
\def\eeqan{\end{eqnarray}}

\let\bar=\overbar

\def\Dslash{\not{\hbox{\kern-4pt $D$}}}
\def\dslash{\not{\hbox{\kern-2pt $\del$}}}

\def\msb{{\bar{\ssstyle M \kern -1pt S}}}

\def\Title#1{\begin{center} {\Large {\bf #1} } \end{center}}

\begin{document}

\Title{Penguin contributions to $\boldsymbol{B\to J/\psi P}$ Decays\footnote
{Proceedings of CKM 2012, the 7th International Workshop on the CKM Unitarity Triangle, University of Cincinnati, USA, 28 September - 2 October 2012.}}

\bigskip\bigskip


\begin{raggedright}  

{\it Martin Jung\index{Jung, M.}\\
Theoretical Physics T IV\\
Technical University Dortmund\\
D-44221 Dortmund, GERMANY}
\bigskip\bigskip
\end{raggedright}

\subsection*{Abstract}
The high precision to which the standard model has been confirmed implies that new physics effects have to be small in the observed processes. Together with the outstanding precision expected from present and future collider experiments this renders the evaluation of subleading SM contributions necessary. For the ``golden mode'', $B_d\to J/\psi K$, these so-called ``penguin pollution'' terms can be controlled by using flavour symmetry relations. A recent analysis is presented which yields a stronger bound on the maximal penguin influence than
previous ones and shows how the corresponding uncertainty can be reduced with coming data.

\section{Introduction}
Roughly 40 years after its proposal \cite{Kobayashi:1973fv}, the Kobayashi-Maskawa mechanism continues to give a consistent interpretation of the available data on flavour observables and CP violation. This fact is reflected in successful fits to the Unitarity Triangle (UT) \cite{UTfits},
where, despite the precision data which have become available during the last decade, still no clear sign of physics beyond the Standard Model (SM) is seen. On the contrary, some of the tensions that used to be present, e.g. in the extraction of the CKM angle $\beta$ ($\phi_1$) \cite{Btaunulit} or $B_s$ mixing (see e.g. \cite{Lenz:2010gu}), have recently been rendered less severe by new data \cite{newdata}.\footnote{The very recent measurement \cite{Aaij:2012ke} has not been included yet in the analysis.}

Other puzzles related to the UT analysis remain, like the difference between $|V_{ub}|$ extracted from inclusive and exclusive semileptonic decays or the largish CP violation in the kaon system, but are less significant. The important lesson from these observations is that new physics (NP) effects in the related observables have to be small. This observation, together with the present and coming experimental improvements at high-luminosity machines, renders precision predictions for the involved observables particularly important; among these, a refined analysis for the extraction of the angle $\beta$ from the ``golden mode'' $B\to J/\psi K_S$ becomes necessary, which constitutes a key measurement in this context. In the following we report the recent proposal for such a refinement \cite{Jung:2012mp}, allowing for continued improvement with coming data.

\section{Penguin Pollution in $\boldsymbol{B\to J/\psi K}$}

The impressive precision obtained for the angle $\beta$ became possible due to the fact that in the ``golden mode'', $B_d\to J/\psi K_S$, explicit calculation of the relevant matrix elements can be avoided once subleading doubly Cabbibo suppressed terms are assumed to vanish \cite{Bigi:1981qs}, in combination with an experimentally accessible final state. However, given the apparent smallness of NP effects and the precision the LHC experiments and the planned next-generation $B$ factory are aiming at for this and related modes, a critical reconsideration of the assumptions used is mandatory. Estimates yield corrections to the famous relation $S_{J/\psi K_S}=\sin\phi_d$ of the order $\mathcal{O}(10^{-3})$, only \cite{PenEstimates}; 
it is, however, notoriously difficult to actually calculate the relevant matrix elements, and non-perturbative enhancements cannot a priori be excluded.

To include these subleading contributions, the size of their matrix elements relative to the leading one has to be determined. An explicit calculation still does not seem feasible to an acceptable precision for the decays in question, as our available tools do not work for these modes; QCD factorization for example does hold formally, but corrections to the heavy quark limit are large, $\mathcal{O}(\Lambda_{QCD}/(\alpha_sm_c))$, which is why it fails to yield a quantitative description, see e.g. \cite{Beneke:2000ry}. This is why typically flavour symmetry relations are used, i.e. $SU(3)$, relating up, down and strange quarks, or its subgroup $U$-spin, including only down and strange quarks. These allow for accessing the unknown matrix element ratios via decays where their relative influence is larger (``control modes'') \cite{PenUspin}. 
This method has the advantage of being completely data-driven, and the resulting value for the $B$ mixing phase provides improved access to NP in mixing once the SM value of this phase is determined independently. 

The main limitations of that approach were firstly the limited data for the control modes, as their rates are suppressed by $\lambda^2\sim5\%$ compared to the one of $B\to J/\psi K$, and secondly unknown corrections to the symmetry limit. The first issue was already rendered less severe by recent data from CDF and LHCb \cite{NewDataJPsiP},
and will be resolved by the LHC in combination with the planned Super Flavour Factory (SFF). The second was addressed in \cite{Jung:2012mp}, as we will explain in the following. The main idea is to include the symmetry-breaking corrections in a model-independent manner on a group-theoretical basis (for earlier applications of this method, see e.g. \cite{SU3breakingmodind}). 
Extending furthermore the symmetry group from $U$-spin (used in \cite{PenUspin}) 
to full $SU(3)$ allows to relate a sufficiently large number of decay modes (the full set of $B\to J/\psi P$ modes, with $B\in\{B_u,B_d,B_s\}$ and $P\in \{\pi^+,\pi^0,K^+,K^0,\bar{K}^0\}$) to determine the $SU(3)$ breaking parameters as well as the penguin pollution ones from the fit, using mild assumptions which are mostly testable with data \cite{Jung:2012mp}.

Applying this method to presently available data for these decays \cite{NewDataJPsiP,HFAGPDG} 
shows clearly the importance of $SU(3)$-breaking effects. Even when allowing for huge values of the penguin parameters, the fit in the $SU(3)$ limit yields $\chi^2_{\rm min}/{\rm d.o.f.}=22.3(23.9)/5$, where the first number corresponds to using the former world average for the ratio $BR(B^-\to J/\psi \pi^-)/BR(B^-\to J/\psi K^-)$ (``dataset 1''), and the second to the new LHCb result (``dataset 2''). The latter yields a value about 3 standard deviations away from the former, which is why the results are compared explicitly instead of averaging the input measurements.
Importantly, correlations to the measured branching ratios drive the shift $\Delta S=-S(B\to J/\psi K_S)+\sin\phi_d$ to relatively large values in this case, in the opposite direction of the tension observed in the UT fit.
It is furthermore interesting to note that the inclusion of neglected contributions of order $\Lambda_{QCD}/m_b$ does not improve the fit, confirming our choice to set them to zero, see also the discussion in \cite{Jung:2012mp}.

In a next step, $SU(3)$-breaking contributions are included in the fit, while neglecting penguin pollution. This fit works rather well, yielding for the two datasets  $\chi^2_{\rm min}=9.4(6.0)$ for 7 effective degrees of freedom\footnote{\emph{Effective degrees of freedom} are defined here as number of observables minus the number of parameters which are effectively changing the fit, see also \cite{Jung:2012mp}.}.
The best fit point yields a ratio of the larger $SU(3)$-breaking matrix element with the leading one of $19(24)\%$, which is perfectly within the expected range for this quantity. Therefore the data at present can be described with the expected amount of $SU(3)$ breaking and negligible penguin contributions.
The parameter point corresponding to factorizable $SU(3)$ breaking lies outside the $95\%$~CL area.  
This is in accordance with \cite{Fleischer:proc}, where, using factorizable $SU(3)$ breaking, a preference for sizable penguin contributions is observed.

In the following, we do not assume factorizable $SU(3)$ breaking and perform the full fit with both additional contributions, i.e. $SU(3)$ breaking and penguin pollution. The fit improves again slightly, to $\chi^2_{\rm min}=2.8(2.3)$ for 3 effective degrees of freedom, when we refrain from applying strong restrictions on the parameter values\footnote{We do not allow for ``exchanging roles'' though, i.e. we continue to assume the leading matrix element to be the one in the $SU(3)$ limit with no penguin contributions.}. In this fit, the $SU(3)$-breaking parameters allow to accommodate the pattern of branching ratios, while the penguin contributions are mainly determined by the CP and isospin asymmetries. The central values of the penguin parameters still tend to larger values than theoretically expected. This is not surprising, given the fact that the isospin asymmetry in $B\to J/\psi K$ has a central value about ten times larger than the naive expectation, however with large uncertainties. The corresponding branching ratios are predicted to be around one standard deviation higher (lower) for $\bar{B}^0\to J/\psi \bar{K}^0\,(B^-\to J/\psi K^-)$, making an  additional measurement of their ratio important, which correspondingly is predicted to take a significantly different central value than the one presently measured.

Restricting the fit parameters  to the expected ranges, i.e. at most an $SU(3)$ breaking of $r_{SU(3)}=40\%$, and a ratio of the penguin matrix element with the leading one of $r_{\rm pen}=50\%$, shows a preference for dataset 2, where the minimal $\chi^2$ remains basically unchanged, while for dataset~1 it doubles approximately. The new result for $BR(B^-\to J/\psi\pi^-)/BR(B^-\to J/\psi K^-)$ obtained by LHCb seems therefore favoured by this fit. While  it is too early to draw conclusions, this observation demonstrates once more the importance of precise branching ratio measurements in this context.
For both datasets, the shift $\Delta S$ now tends again to positive values, thereby lowering the corresponding tension in the UT fit. It is however still compatible with zero, in agreement with the above observation of a reasonable fit without penguin terms. The obtained ranges read
\begin{eqnarray}
\Delta S_{J/\psi K}^{\rm set\,1} &=& [0.001,0.005] ([-0.004,0.011])\,,\mbox{\quad and }\\
\Delta S_{J/\psi K}^{\rm set\,2} &=& [0.004,0.011] ([-0.003,0.012])\,,
\end{eqnarray}
for $68\%$ ($95\%$)~CL, respectively, where the preferred sign change compared to the $SU(3)$ limit is due to relaxed correlations between $S(B\to J/\psi \pi^0)$ and the branching ratios in the fit, because of the additional contributions. This underlines the necessity to treat $SU(3)$ breaking model-independently.
Note that $S(B_d\to J/\psi\pi^0)$ is predicted to lie below the present central value of the measurement, thereby supporting the Belle result \cite{Lee:2007wd} over the BaBar one \cite{Aubert:2008bs}, which indicates a very large value for this observable. 
\begin{figure}[hbt]
\begin{center}
\includegraphics[width=7.2cm]{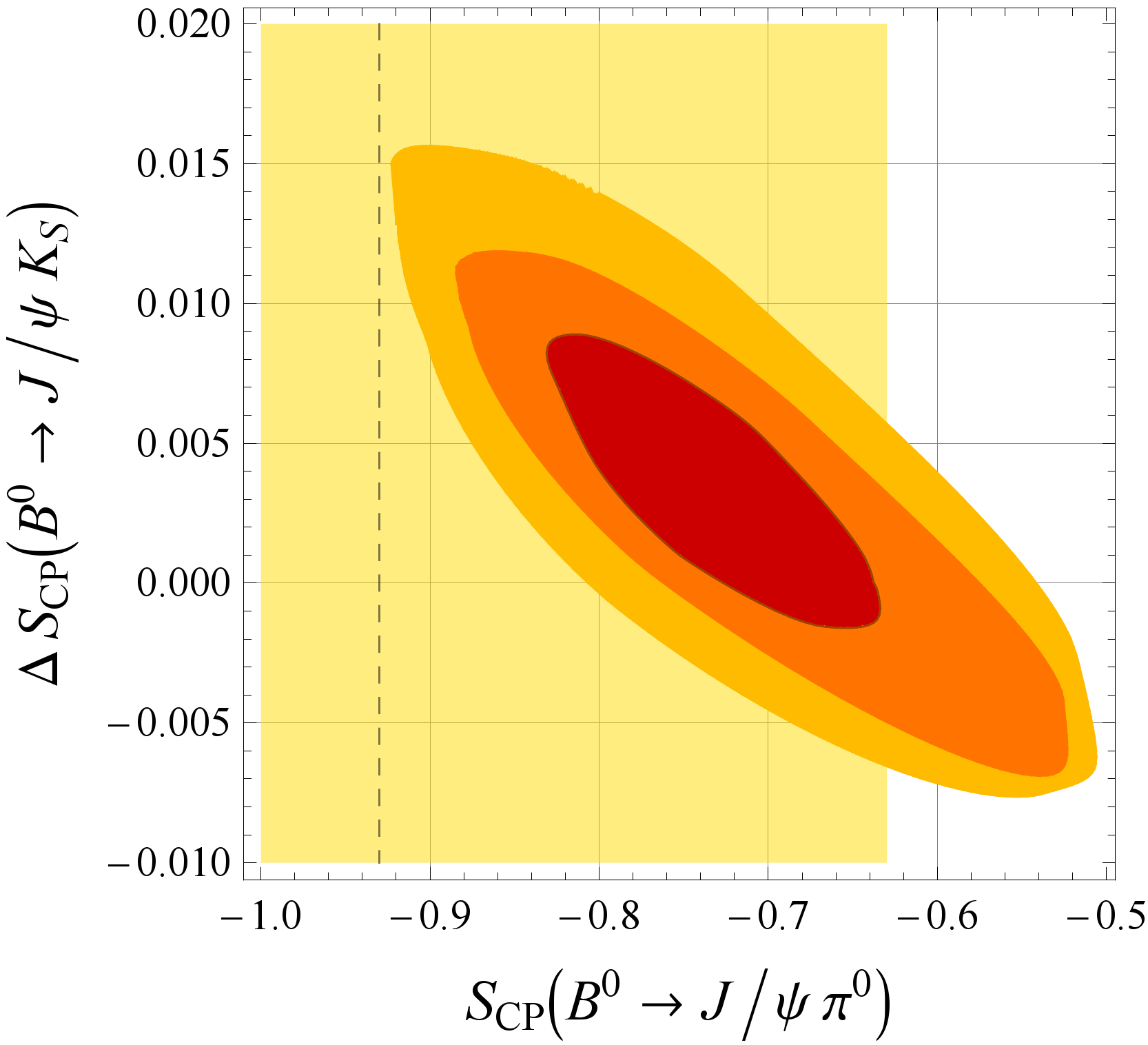}\hfill \includegraphics[width=7.2cm]{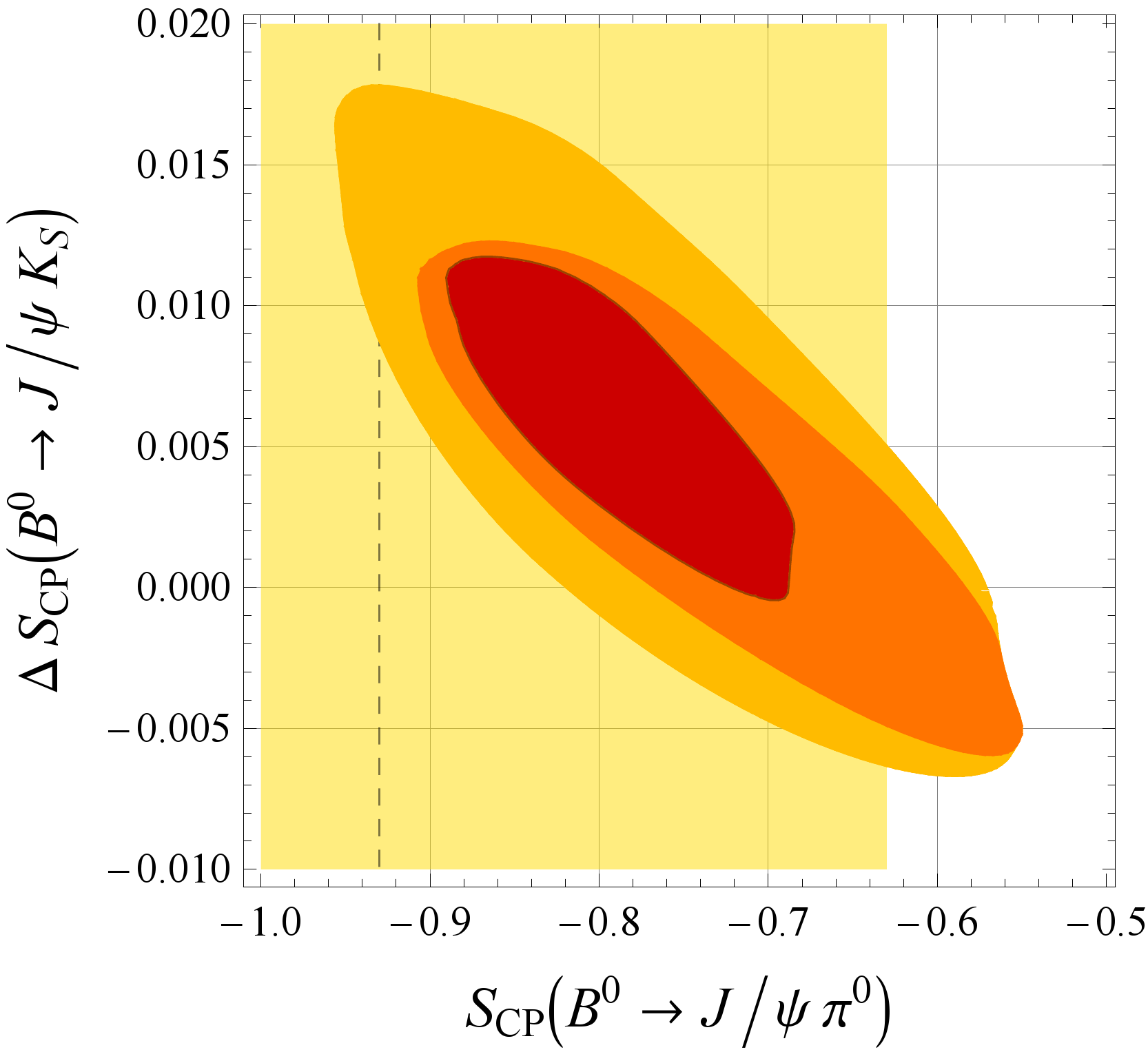}
\caption{\label{fig::resultsfullfita} Fit results for datasets 1 (left) and 2 (right), for $\Delta S$ versus $S_{\rm CP}(B^0\to J/\psi \pi^0)$, including all available data. The inner areas correspond to $68\%$ CL and $95\%$ CL with $r_{SU(3)}=40\%$ and $r_{\rm pen}=50\%$. The outer one is shown for illustration purposes, only, and corresponds to $95\%$ CL when allowing for up to $r_{SU(3)}=60\%$ and $r_{\rm pen}=75\%$. The light yellow area indicates the 2-$\sigma$ range of the $S(B^0\to J/\psi \pi^0)$ average, the dashed line its central value. Figure taken from  \cite{Jung:2012mp}.}
\end{center}
\end{figure}
These findings are illustrated in Fig.~\ref{fig::resultsfullfita}.
The same fit allows to predict the so far unmeasured CP asymmetries in $B_s\to J/\psi K$ decays: their absolute values lie for both datasets below approximately $30\%$ at $95\%$~CL. On the one hand this allows for a crosscheck for the description in the above framework. On the other hand it is clear that a measurement with a precision of $\sim10\%$ will already yield a significant additional constraint on the model parameters. Especially the dependence on the (already weak) theory assumptions will be further reduced with such a measurement \cite{Jung:2012mp}.

The mixing phase is extracted as $\phi_d^{\rm fit}=0.74\pm0.03$ (equal for both datasets), which is to be compared with $\phi_{d,{\rm naive}}^{\rm SM}=0.73\pm0.03$ when using the naive relation without penguin contributions. The inclusion of the correction therefore yields the same precision, but induces a shift of the central value. The same is true for future data, as shown in \cite{Jung:2012mp} by the consideration of several scenarios corresponding to additional data from the LHCb and SFF experiments. This implies the corresponding error to be reducible, and therefore ensures the golden mode to keep its special position among flavour observables.

The comparison of the full fit with ones using the assumption of factorizable $SU(3)$ breaking is complicated by the fact that this assumption leads to several relations among the parameters used here, implying very different numbers of degrees of freedom. A more detailed discussion is therefore left for future work. However, generically the resulting predictions for CP asymmetries should be larger when assuming factorizable breaking, allowing this subject to be clarified experimentally.

In principle, the same approach can be used to constrain penguin pollution in the other ``golden mode'', $B_s\to J/\psi\phi$. Technical difficulties are the fact that the $\phi$ meson does not belong to a single representation, and the more complicated structure of the final state. The latter is also complicating the experimental analysis; however, recently progress has been made in measuring additional modes \cite{newdataBtoJPsiV}. If the $b\to d$ modes can be measured sufficiently precise to control the penguin pollution as well as the $SU(3)$ breaking is subject to further studies.

\section{Conclusions}
CP violation studies in heavy meson systems remain a very active field, and one of the main paths to discover NP. The general picture remains consistent with the KM mechanism as the only source of low-energy CP violation; in fact, the fits have improved very recently due to a new measurement for $B\to\tau\nu$. 

This -- in many ways unexpected -- situation requires a more precise knowledge of the corresponding SM predictions, as potential small NP contributions will compete with subleading SM ones. The ``golden mode'' $B_d\to J/\psi K$ is an example where subleading contributions can affect the extraction of the mixing phase. For this mode a new approach to control them has been advocated, allowing to take into account $SU(3)$ corrections model-independently, which were shown to affect the procedure severely. The main result is a new limit, $|\Delta S_{J/\psi K}|\lesssim 0.01$ ($95\%$~CL), which can additionally be improved by coming data.

In conclusion, the apparent smallness of NP effects in flavour observables poses a challenge to both theory and experiment. On the experimental side it is met by several high-luminosity collider experiments, both running and under construction, allowing for unprecedented precision. Also on the theory side the challenge is answered, by new strategies and adapting known ones to higher precision. Together, these developments make for an exciting way ahead.

\section*{Acknowledgements}
I want to thank the organizers for the invitation and an enjoyable workshop.
This work is supported by the Bundesministerium f\"ur Bildung und Forschung (BMBF).


\begin{thebibliography}{32}%
\makeatletter
\providecommand \@ifxundefined [1]{%
 \@ifx{#1\undefined}
}%
\providecommand \@ifnum [1]{%
 \ifnum #1\expandafter \@firstoftwo
 \else \expandafter \@secondoftwo
 \fi
}%
\providecommand \@ifx [1]{%
 \ifx #1\expandafter \@firstoftwo
 \else \expandafter \@secondoftwo
 \fi
}%
\providecommand \natexlab [1]{#1}%
\providecommand \enquote  [1]{``#1''}%
\providecommand \bibnamefont  [1]{#1}%
\providecommand \bibfnamefont [1]{#1}%
\providecommand \citenamefont [1]{#1}%
\providecommand \href@noop [0]{\@secondoftwo}%
\providecommand \href [0]{\begingroup \@sanitize@url \@href}%
\providecommand \@href[1]{\@@startlink{#1}\@@href}%
\providecommand \@@href[1]{\endgroup#1\@@endlink}%
\providecommand \@sanitize@url [0]{\catcode `\\12\catcode `\$12\catcode
  `\&12\catcode `\#12\catcode `\^12\catcode `\_12\catcode `\%12\relax}%
\providecommand \@@startlink[1]{}%
\providecommand \@@endlink[0]{}%
\providecommand \url  [0]{\begingroup\@sanitize@url \@url }%
\providecommand \@url [1]{\endgroup\@href {#1}{\urlprefix }}%
\providecommand \urlprefix  [0]{URL }%
\providecommand \Eprint [0]{\href }%
\providecommand \doibase [0]{http://dx.doi.org/}%
\providecommand \selectlanguage [0]{\@gobble}%
\providecommand \bibinfo  [0]{\@secondoftwo}%
\providecommand \bibfield  [0]{\@secondoftwo}%
\providecommand \translation [1]{[#1]}%
\providecommand \BibitemOpen [0]{}%
\providecommand \bibitemStop [0]{}%
\providecommand \bibitemNoStop [0]{.\EOS\space}%
\providecommand \EOS [0]{\spacefactor3000\relax}%
\providecommand \BibitemShut  [1]{\csname bibitem#1\endcsname}%
\let\auto@bib@innerbib\@empty
\bibitem [{\citenamefont {Jung}(2012{\natexlab{a}})}]{Jung:2012pz}%
  \BibitemOpen
  \bibfield  {author} {\bibinfo {author} {\bibfnamefont {M.}~\bibnamefont
  {Jung}},\ }\href@noop {} {\bibfield  {journal} {\bibinfo  {journal} {PoS}\
  }\textbf {\bibinfo {volume} {HQL2012}},\ \bibinfo {pages} {037} (\bibinfo
  {year} {2012}{\natexlab{a}})},\ \Eprint {http://arxiv.org/abs/1208.1286}
  {arXiv:1208.1286 [hep-ph]} \BibitemShut {NoStop}%
\bibitem [{\citenamefont {Kobayashi}\ and\ \citenamefont
  {Maskawa}(1973)}]{Kobayashi:1973fv}%
  \BibitemOpen
  \bibfield  {author} {\bibinfo {author} {\bibfnamefont {M.}~\bibnamefont
  {Kobayashi}}\ and\ \bibinfo {author} {\bibfnamefont {T.}~\bibnamefont
  {Maskawa}},\ }\href@noop {} {\bibfield  {journal} {\bibinfo  {journal} {Prog.
  Theor. Phys.}\ }\textbf {\bibinfo {volume} {49}},\ \bibinfo {pages} {652}
  (\bibinfo {year} {1973})}\BibitemShut {NoStop}%
\bibitem [{\citenamefont {Charles}\ \emph {et~al.}(2005)\citenamefont {Charles}
  \emph {et~al.}}]{Charles:2004jd}%
  \BibitemOpen
  \bibfield  {author} {\bibinfo {author} {\bibfnamefont {J.}~\bibnamefont
  {Charles}} \emph {et~al.} (\bibinfo {collaboration} {CKMfitter Group}),\
  }\href@noop {} {\bibfield  {journal} {\bibinfo  {journal} {Eur. Phys. J.}\
  }\textbf {\bibinfo {volume} {C41}},\ \bibinfo {pages} {1} (\bibinfo {year}
  {2005})},\ \bibinfo {note} {updated results and plots available at: {\tt
  http://ckmfitter.in2p3.fr}},\ \Eprint {http://arxiv.org/abs/hep-ph/0406184}
  {hep-ph/0406184} \BibitemShut {NoStop}%
\bibitem [{\citenamefont {Ciuchini}\ \emph {et~al.}(2001)\citenamefont
  {Ciuchini} \emph {et~al.}}]{Ciuchini:2000de}%
  \BibitemOpen
  \bibfield  {author} {\bibinfo {author} {\bibfnamefont {M.}~\bibnamefont
  {Ciuchini}} \emph {et~al.},\ }\href@noop {} {\bibfield  {journal} {\bibinfo
  {journal} {JHEP}\ }\textbf {\bibinfo {volume} {07}},\ \bibinfo {pages} {013}
  (\bibinfo {year} {2001})},\ \bibinfo {note} {updated results and plots
  available at: {\tt http://www.utfit.org}},\ \Eprint
  {http://arxiv.org/abs/hep-ph/0012308} {hep-ph/0012308} \BibitemShut {NoStop}%
\bibitem [{\citenamefont {Deschamps}(2008)}]{Deschamps:2008de}%
  \BibitemOpen
  \bibfield  {author} {\bibinfo {author} {\bibfnamefont {O.}~\bibnamefont
  {Deschamps}},\ }\href@noop {} {\  (\bibinfo {year} {2008})},\ \Eprint
  {http://arxiv.org/abs/0810.3139} {arXiv:0810.3139 [hep-ph]} \BibitemShut
  {NoStop}%
\bibitem [{\citenamefont {Feldmann}\ \emph {et~al.}(2008)\citenamefont
  {Feldmann}, \citenamefont {Jung},\ and\ \citenamefont
  {Mannel}}]{Feldmann:2008fb}%
  \BibitemOpen
  \bibfield  {author} {\bibinfo {author} {\bibfnamefont {T.}~\bibnamefont
  {Feldmann}}, \bibinfo {author} {\bibfnamefont {M.}~\bibnamefont {Jung}}, \
  and\ \bibinfo {author} {\bibfnamefont {T.}~\bibnamefont {Mannel}},\
  }\href@noop {} {\bibfield  {journal} {\bibinfo  {journal} {JHEP}\ }\textbf
  {\bibinfo {volume} {08}},\ \bibinfo {pages} {066} (\bibinfo {year} {2008})},\
  \Eprint {http://arxiv.org/abs/0803.3729 (hep-ph)} {0803.3729 (hep-ph)}
  \BibitemShut {NoStop}%
\bibitem [{\citenamefont {Bona}\ \emph {et~al.}(2010)\citenamefont {Bona} \emph
  {et~al.}}]{Bona:2009cj}%
  \BibitemOpen
  \bibfield  {author} {\bibinfo {author} {\bibfnamefont {M.}~\bibnamefont
  {Bona}} \emph {et~al.} (\bibinfo {collaboration} {UTfit Collaboration}),\
  }\href {\doibase 10.1016/j.physletb.2010.02.063} {\bibfield  {journal}
  {\bibinfo  {journal} {Phys.Lett.}\ }\textbf {\bibinfo {volume} {B687}},\
  \bibinfo {pages} {61} (\bibinfo {year} {2010})},\ \Eprint
  {http://arxiv.org/abs/0908.3470} {arXiv:0908.3470 [hep-ph]} \BibitemShut
  {NoStop}%
\bibitem [{\citenamefont {Lunghi}\ and\ \citenamefont
  {Soni}(2010)}]{Lunghi:2009ke}%
  \BibitemOpen
  \bibfield  {author} {\bibinfo {author} {\bibfnamefont {E.}~\bibnamefont
  {Lunghi}}\ and\ \bibinfo {author} {\bibfnamefont {A.}~\bibnamefont {Soni}},\
  }\href {\doibase 10.1103/PhysRevLett.104.251802} {\bibfield  {journal}
  {\bibinfo  {journal} {Phys.Rev.Lett.}\ }\textbf {\bibinfo {volume} {104}},\
  \bibinfo {pages} {251802} (\bibinfo {year} {2010})},\ \Eprint
  {http://arxiv.org/abs/0912.0002} {arXiv:0912.0002 [hep-ph]} \BibitemShut
  {NoStop}%
\bibitem [{\citenamefont {Charles}\ \emph {et~al.}(2011)\citenamefont
  {Charles}, \citenamefont {Deschamps}, \citenamefont {Descotes-Genon},
  \citenamefont {Itoh}, \citenamefont {Lacker} \emph
  {et~al.}}]{Charles:2011va}%
  \BibitemOpen
  \bibfield  {author} {\bibinfo {author} {\bibfnamefont {J.}~\bibnamefont
  {Charles}}, \bibinfo {author} {\bibfnamefont {O.}~\bibnamefont {Deschamps}},
  \bibinfo {author} {\bibfnamefont {S.}~\bibnamefont {Descotes-Genon}},
  \bibinfo {author} {\bibfnamefont {R.}~\bibnamefont {Itoh}}, \bibinfo {author}
  {\bibfnamefont {H.}~\bibnamefont {Lacker}},  \emph {et~al.},\ }\href@noop {}
  {\bibfield  {journal} {\bibinfo  {journal} {Phys.Rev.}\ }\textbf {\bibinfo
  {volume} {D84}},\ \bibinfo {pages} {033005} (\bibinfo {year} {2011})},\
  \Eprint {http://arxiv.org/abs/1106.4041} {arXiv:1106.4041 [hep-ph]}
  \BibitemShut {NoStop}%
\bibitem [{\citenamefont {Adachi}\ \emph {et~al.}(2012)\citenamefont {Adachi}
  \emph {et~al.}}]{Adachi:2012mm}%
  \BibitemOpen
  \bibfield  {author} {\bibinfo {author} {\bibfnamefont {I.}~\bibnamefont
  {Adachi}} \emph {et~al.} (\bibinfo {collaboration} {Belle Collaboration}),\
  }\href@noop {} {\  (\bibinfo {year} {2012})},\ \Eprint
  {http://arxiv.org/abs/1208.4678} {arXiv:1208.4678 [hep-ex]} \BibitemShut
  {NoStop}%
\bibitem [{\citenamefont {Bigi}\ and\ \citenamefont
  {Sanda}(1981)}]{Bigi:1981qs}%
  \BibitemOpen
  \bibfield  {author} {\bibinfo {author} {\bibfnamefont {I.~I.}\ \bibnamefont
  {Bigi}}\ and\ \bibinfo {author} {\bibfnamefont {A.}~\bibnamefont {Sanda}},\
  }\href {\doibase 10.1016/0550-3213(81)90519-8} {\bibfield  {journal}
  {\bibinfo  {journal} {Nucl.Phys.}\ }\textbf {\bibinfo {volume} {B193}},\
  \bibinfo {pages} {85} (\bibinfo {year} {1981})},\ \bibinfo {note} {dedicated
  to Y. Orloff}\BibitemShut {NoStop}%
\bibitem [{\citenamefont {Boos}\ \emph {et~al.}(2004)\citenamefont {Boos},
  \citenamefont {Mannel},\ and\ \citenamefont {Reuter}}]{Boos:2004xp}%
  \BibitemOpen
  \bibfield  {author} {\bibinfo {author} {\bibfnamefont {H.}~\bibnamefont
  {Boos}}, \bibinfo {author} {\bibfnamefont {T.}~\bibnamefont {Mannel}}, \ and\
  \bibinfo {author} {\bibfnamefont {J.}~\bibnamefont {Reuter}},\ }\href@noop {}
  {\bibfield  {journal} {\bibinfo  {journal} {Phys. Rev.}\ }\textbf {\bibinfo
  {volume} {D70}},\ \bibinfo {pages} {036006} (\bibinfo {year} {2004})},\
  \Eprint {http://arxiv.org/abs/hep-ph/0403085} {hep-ph/0403085} \BibitemShut
  {NoStop}%
\bibitem [{\citenamefont {Li}\ and\ \citenamefont {Mishima}(2007)}]{Li:2006vq}%
  \BibitemOpen
  \bibfield  {author} {\bibinfo {author} {\bibfnamefont {H.-n.}\ \bibnamefont
  {Li}}\ and\ \bibinfo {author} {\bibfnamefont {S.}~\bibnamefont {Mishima}},\
  }\href@noop {} {\bibfield  {journal} {\bibinfo  {journal} {JHEP}\ }\textbf
  {\bibinfo {volume} {03}},\ \bibinfo {pages} {009} (\bibinfo {year} {2007})},\
  \Eprint {http://arxiv.org/abs/hep-ph/0610120} {hep-ph/0610120} \BibitemShut
  {NoStop}%
\bibitem [{\citenamefont {Gronau}\ and\ \citenamefont
  {Rosner}(2009)}]{Gronau:2008cc}%
  \BibitemOpen
  \bibfield  {author} {\bibinfo {author} {\bibfnamefont {M.}~\bibnamefont
  {Gronau}}\ and\ \bibinfo {author} {\bibfnamefont {J.~L.}\ \bibnamefont
  {Rosner}},\ }\href@noop {} {\bibfield  {journal} {\bibinfo  {journal} {Phys.
  Lett.}\ }\textbf {\bibinfo {volume} {B672}},\ \bibinfo {pages} {349}
  (\bibinfo {year} {2009})},\ \Eprint {http://arxiv.org/abs/0812.4796 (hep-ph)}
  {0812.4796 (hep-ph)} \BibitemShut {NoStop}%
\bibitem [{\citenamefont {Lenz}(2011)}]{Lenz:2011zz}%
  \BibitemOpen
  \bibfield  {author} {\bibinfo {author} {\bibfnamefont {A.~J.}\ \bibnamefont
  {Lenz}},\ }\href@noop {} {\bibfield  {journal} {\bibinfo  {journal}
  {Phys.Rev.}\ }\textbf {\bibinfo {volume} {D84}},\ \bibinfo {pages} {031501}
  (\bibinfo {year} {2011})},\ \Eprint {http://arxiv.org/abs/1106.3200}
  {arXiv:1106.3200 [hep-ph]} \BibitemShut {NoStop}%
\bibitem [{\citenamefont {Fleischer}(1999)}]{Fleischer:1999nz}%
  \BibitemOpen
  \bibfield  {author} {\bibinfo {author} {\bibfnamefont {R.}~\bibnamefont
  {Fleischer}},\ }\href@noop {} {\bibfield  {journal} {\bibinfo  {journal}
  {Phys. Lett.}\ }\textbf {\bibinfo {volume} {B459}},\ \bibinfo {pages} {306}
  (\bibinfo {year} {1999})},\ \Eprint {http://arxiv.org/abs/hep-ph/9903456}
  {hep-ph/9903456} \BibitemShut {NoStop}%
\bibitem [{\citenamefont {Ciuchini}\ \emph {et~al.}(2005)\citenamefont
  {Ciuchini}, \citenamefont {Pierini},\ and\ \citenamefont
  {Silvestrini}}]{Ciuchini:2005mg}%
  \BibitemOpen
  \bibfield  {author} {\bibinfo {author} {\bibfnamefont {M.}~\bibnamefont
  {Ciuchini}}, \bibinfo {author} {\bibfnamefont {M.}~\bibnamefont {Pierini}}, \
  and\ \bibinfo {author} {\bibfnamefont {L.}~\bibnamefont {Silvestrini}},\
  }\href@noop {} {\bibfield  {journal} {\bibinfo  {journal} {Phys. Rev. Lett.}\
  }\textbf {\bibinfo {volume} {95}},\ \bibinfo {pages} {221804} (\bibinfo
  {year} {2005})},\ \Eprint {http://arxiv.org/abs/hep-ph/0507290}
  {hep-ph/0507290} \BibitemShut {NoStop}%
\bibitem [{\citenamefont {Ciuchini}\ \emph {et~al.}(2011)\citenamefont
  {Ciuchini}, \citenamefont {Pierini},\ and\ \citenamefont
  {Silvestrini}}]{Ciuchini:2011kd}%
  \BibitemOpen
  \bibfield  {author} {\bibinfo {author} {\bibfnamefont {M.}~\bibnamefont
  {Ciuchini}}, \bibinfo {author} {\bibfnamefont {M.}~\bibnamefont {Pierini}}, \
  and\ \bibinfo {author} {\bibfnamefont {L.}~\bibnamefont {Silvestrini}},\
  }\href@noop {} {\  (\bibinfo {year} {2011})},\ \Eprint
  {http://arxiv.org/abs/1102.0392} {arXiv:1102.0392 [hep-ph]} \BibitemShut
  {NoStop}%
\bibitem [{\citenamefont {Faller}\ \emph
  {et~al.}(2009{\natexlab{a}})\citenamefont {Faller}, \citenamefont {Jung},
  \citenamefont {Fleischer},\ and\ \citenamefont {Mannel}}]{Faller:2008zc}%
  \BibitemOpen
  \bibfield  {author} {\bibinfo {author} {\bibfnamefont {S.}~\bibnamefont
  {Faller}}, \bibinfo {author} {\bibfnamefont {M.}~\bibnamefont {Jung}},
  \bibinfo {author} {\bibfnamefont {R.}~\bibnamefont {Fleischer}}, \ and\
  \bibinfo {author} {\bibfnamefont {T.}~\bibnamefont {Mannel}},\ }\href@noop {}
  {\bibfield  {journal} {\bibinfo  {journal} {Phys. Rev.}\ }\textbf {\bibinfo
  {volume} {D79}},\ \bibinfo {pages} {014030} (\bibinfo {year}
  {2009}{\natexlab{a}})},\ \Eprint {http://arxiv.org/abs/0809.0842 (hep-ph)}
  {0809.0842 (hep-ph)} \BibitemShut {NoStop}%
\bibitem [{\citenamefont {Faller}\ \emph
  {et~al.}(2009{\natexlab{b}})\citenamefont {Faller}, \citenamefont
  {Fleischer},\ and\ \citenamefont {Mannel}}]{Faller:2008gt}%
  \BibitemOpen
  \bibfield  {author} {\bibinfo {author} {\bibfnamefont {S.}~\bibnamefont
  {Faller}}, \bibinfo {author} {\bibfnamefont {R.}~\bibnamefont {Fleischer}}, \
  and\ \bibinfo {author} {\bibfnamefont {T.}~\bibnamefont {Mannel}},\ }\href
  {\doibase 10.1103/PhysRevD.79.014005} {\bibfield  {journal} {\bibinfo
  {journal} {Phys.Rev.}\ }\textbf {\bibinfo {volume} {D79}},\ \bibinfo {pages}
  {014005} (\bibinfo {year} {2009}{\natexlab{b}})},\ \Eprint
  {http://arxiv.org/abs/0810.4248} {arXiv:0810.4248 [hep-ph]} \BibitemShut
  {NoStop}%
\bibitem [{\citenamefont {Aaltonen}\ \emph {et~al.}(2011)\citenamefont
  {Aaltonen} \emph {et~al.}}]{Aaltonen:2011sy}%
  \BibitemOpen
  \bibfield  {author} {\bibinfo {author} {\bibfnamefont {T.}~\bibnamefont
  {Aaltonen}} \emph {et~al.} (\bibinfo {collaboration} {CDF Collaboration}),\
  }\href {\doibase 10.1103/PhysRevD.83.052012} {\bibfield  {journal} {\bibinfo
  {journal} {Phys.Rev.}\ }\textbf {\bibinfo {volume} {D83}},\ \bibinfo {pages}
  {052012} (\bibinfo {year} {2011})},\ \Eprint {http://arxiv.org/abs/1102.1961}
  {arXiv:1102.1961 [hep-ex]} \BibitemShut {NoStop}%
\bibitem [{\citenamefont {Aaij}\ \emph
  {et~al.}(2012{\natexlab{a}})\citenamefont {Aaij} \emph
  {et~al.}}]{Aaij:2012di}%
  \BibitemOpen
  \bibfield  {author} {\bibinfo {author} {\bibfnamefont {R.}~\bibnamefont
  {Aaij}} \emph {et~al.} (\bibinfo {collaboration} {LHCb Collaboration}),\
  }\href {\doibase 10.1016/j.physletb.2012.05.062} {\bibfield  {journal}
  {\bibinfo  {journal} {Phys.Lett.}\ }\textbf {\bibinfo {volume} {B713}},\
  \bibinfo {pages} {172} (\bibinfo {year} {2012}{\natexlab{a}})},\ \Eprint
  {http://arxiv.org/abs/1205.0934} {arXiv:1205.0934 [hep-ex]} \BibitemShut
  {NoStop}%
\bibitem [{\citenamefont {Aaij}\ \emph
  {et~al.}(2012{\natexlab{b}})\citenamefont {Aaij} \emph
  {et~al.}}]{Aaij:2012jw}%
  \BibitemOpen
  \bibfield  {author} {\bibinfo {author} {\bibfnamefont {R.}~\bibnamefont
  {Aaij}} \emph {et~al.} (\bibinfo {collaboration} {LHCb collaboration}),\
  }\href {\doibase 10.1103/PhysRevD.85.091105} {\bibfield  {journal} {\bibinfo
  {journal} {Phys.Rev.}\ }\textbf {\bibinfo {volume} {D85}},\ \bibinfo {pages}
  {091105} (\bibinfo {year} {2012}{\natexlab{b}})},\ \Eprint
  {http://arxiv.org/abs/1203.3592} {arXiv:1203.3592 [hep-ex]} \BibitemShut
  {NoStop}%
\bibitem [{\citenamefont {Jung}(2012{\natexlab{b}})}]{Jung:2012mp}%
  \BibitemOpen
  \bibfield  {author} {\bibinfo {author} {\bibfnamefont {M.}~\bibnamefont
  {Jung}},\ }\href@noop {} {\  (\bibinfo {year} {2012}{\natexlab{b}})},\
  \Eprint {http://arxiv.org/abs/1206.2050} {arXiv:1206.2050 [hep-ph]}
  \BibitemShut {NoStop}%
\bibitem [{\citenamefont {Savage}(1991)}]{Savage:1991wu}%
  \BibitemOpen
  \bibfield  {author} {\bibinfo {author} {\bibfnamefont {M.~J.}\ \bibnamefont
  {Savage}},\ }\href {\doibase 10.1016/0370-2693(91)91917-K} {\bibfield
  {journal} {\bibinfo  {journal} {Phys.Lett.}\ }\textbf {\bibinfo {volume}
  {B257}},\ \bibinfo {pages} {414} (\bibinfo {year} {1991})}\BibitemShut
  {NoStop}%
\bibitem [{\citenamefont {Gronau}\ \emph {et~al.}(1995)\citenamefont {Gronau},
  \citenamefont {Hernandez}, \citenamefont {London},\ and\ \citenamefont
  {Rosner}}]{Gronau:1995hm}%
  \BibitemOpen
  \bibfield  {author} {\bibinfo {author} {\bibfnamefont {M.}~\bibnamefont
  {Gronau}}, \bibinfo {author} {\bibfnamefont {O.~F.}\ \bibnamefont
  {Hernandez}}, \bibinfo {author} {\bibfnamefont {D.}~\bibnamefont {London}}, \
  and\ \bibinfo {author} {\bibfnamefont {J.~L.}\ \bibnamefont {Rosner}},\
  }\href@noop {} {\bibfield  {journal} {\bibinfo  {journal} {Phys. Rev.}\
  }\textbf {\bibinfo {volume} {D52}},\ \bibinfo {pages} {6356} (\bibinfo {year}
  {1995})},\ \Eprint {http://arxiv.org/abs/hep-ph/9504326} {hep-ph/9504326}
  \BibitemShut {NoStop}%
\bibitem [{\citenamefont {Grinstein}\ and\ \citenamefont
  {Lebed}(1996)}]{Grinstein:1996us}%
  \BibitemOpen
  \bibfield  {author} {\bibinfo {author} {\bibfnamefont {B.}~\bibnamefont
  {Grinstein}}\ and\ \bibinfo {author} {\bibfnamefont {R.~F.}\ \bibnamefont
  {Lebed}},\ }\href {\doibase 10.1103/PhysRevD.53.6344} {\bibfield  {journal}
  {\bibinfo  {journal} {Phys.Rev.}\ }\textbf {\bibinfo {volume} {D53}},\
  \bibinfo {pages} {6344} (\bibinfo {year} {1996})},\ \Eprint
  {http://arxiv.org/abs/hep-ph/9602218} {arXiv:hep-ph/9602218 [hep-ph]}
  \BibitemShut {NoStop}%
\bibitem [{\citenamefont {Jung}\ and\ \citenamefont
  {Mannel}(2009)}]{Jung:2009pb}%
  \BibitemOpen
  \bibfield  {author} {\bibinfo {author} {\bibfnamefont {M.}~\bibnamefont
  {Jung}}\ and\ \bibinfo {author} {\bibfnamefont {T.}~\bibnamefont {Mannel}},\
  }\href {\doibase 10.1103/PhysRevD.80.116002} {\bibfield  {journal} {\bibinfo
  {journal} {Phys.Rev.}\ }\textbf {\bibinfo {volume} {D80}},\ \bibinfo {pages}
  {116002} (\bibinfo {year} {2009})},\ \Eprint {http://arxiv.org/abs/0907.0117}
  {arXiv:0907.0117 [hep-ph]} \BibitemShut {NoStop}%
\bibitem [{\citenamefont {Amhis}\ \emph {et~al.}(2012)\citenamefont {Amhis}
  \emph {et~al.}}]{Amhis:2012bh}%
  \BibitemOpen
  \bibfield  {author} {\bibinfo {author} {\bibfnamefont {Y.}~\bibnamefont
  {Amhis}} \emph {et~al.} (\bibinfo {collaboration} {Heavy Flavor Averaging
  Group}),\ }\href@noop {} {\  (\bibinfo {year} {2012})},\ \bibinfo {note}
  {online update available at {\tt http://www.slac.stanford.edu/xorg/hfag}},\
  \Eprint {http://arxiv.org/abs/1207.1158} {arXiv:1207.1158 [hep-ex]}
  \BibitemShut {NoStop}%
\bibitem [{\citenamefont {Beringer}\ \emph {et~al.}(2012)\citenamefont
  {Beringer} \emph {et~al.}}]{Beringer:1900zz}%
  \BibitemOpen
  \bibfield  {author} {\bibinfo {author} {\bibfnamefont {J.}~\bibnamefont
  {Beringer}} \emph {et~al.} (\bibinfo {collaboration} {Particle Data Group}),\
  }\href {\doibase 10.1103/PhysRevD.86.010001} {\bibfield  {journal} {\bibinfo
  {journal} {Phys.Rev.}\ }\textbf {\bibinfo {volume} {D86}},\ \bibinfo {pages}
  {010001} (\bibinfo {year} {2012})}\BibitemShut {NoStop}%
\bibitem [{\citenamefont {Lee}\ \emph {et~al.}(2008)\citenamefont {Lee} \emph
  {et~al.}}]{Lee:2007wd}%
  \BibitemOpen
  \bibfield  {author} {\bibinfo {author} {\bibfnamefont {S.}~\bibnamefont
  {Lee}} \emph {et~al.} (\bibinfo {collaboration} {Belle Collaboration}),\
  }\href {\doibase 10.1103/PhysRevD.77.071101} {\bibfield  {journal} {\bibinfo
  {journal} {Phys.Rev.}\ }\textbf {\bibinfo {volume} {D77}},\ \bibinfo {pages}
  {071101} (\bibinfo {year} {2008})},\ \Eprint {http://arxiv.org/abs/0708.0304}
  {arXiv:0708.0304 [hep-ex]} \BibitemShut {NoStop}%
\bibitem [{\citenamefont {Aubert}\ \emph {et~al.}(2008)\citenamefont {Aubert}
  \emph {et~al.}}]{Aubert:2008bs}%
  \BibitemOpen
  \bibfield  {author} {\bibinfo {author} {\bibfnamefont {B.}~\bibnamefont
  {Aubert}} \emph {et~al.} (\bibinfo {collaboration} {BABAR Collaboration}),\
  }\href {\doibase 10.1103/PhysRevLett.101.021801} {\bibfield  {journal}
  {\bibinfo  {journal} {Phys.Rev.Lett.}\ }\textbf {\bibinfo {volume} {101}},\
  \bibinfo {pages} {021801} (\bibinfo {year} {2008})},\ \Eprint
  {http://arxiv.org/abs/0804.0896} {arXiv:0804.0896 [hep-ex]} \BibitemShut
  {NoStop}%
\end{thebibliography}%


\begin{thebibliography}{99}
\bibitem{Kobayashi:1973fv}
  M.~Kobayashi and T.~Maskawa,
  Prog.\ Theor.\ Phys.\  {\bf 49} (1973) 652.

\bibitem{UTfits}
  J.~Charles {\it et al.}  [CKMfitter Group Collaboration],
  Eur.\ Phys.\ J.\ C {\bf 41} (2005) 1
  [hep-ph/0406184]. Updated results and plots available at: {\tt http://ckmfitter.in2p3.fr}.
%
  M.~Ciuchini, G.~D'Agostini, E.~Franco, V.~Lubicz, G.~Martinelli, F.~Parodi, P.~Roudeau and A.~Stocchi,
  JHEP {\bf 0107} (2001) 013
  [hep-ph/0012308]. Updated results and plots available at: {\tt http://www.utfit.org}.


\bibitem{Btaunulit}
  O.~Deschamps,
  arXiv:0810.3139 [hep-ph].
%
  T.~Feldmann, M.~Jung and T.~Mannel,
  JHEP {\bf 0808} (2008) 066
  [arXiv:0803.3729 [hep-ph]].
%
  M.~Bona {\it et al.}  [UTfit Collaboration],
  Phys.\ Lett.\ B {\bf 687} (2010) 61
  [arXiv:0908.3470 [hep-ph]].
%
  E.~Lunghi and A.~Soni,
  Phys.\ Rev.\ Lett.\  {\bf 104} (2010) 251802
  [arXiv:0912.0002 [hep-ph]].
%
  J.~Charles, O.~Deschamps, S.~Descotes-Genon, R.~Itoh, H.~Lacker, A.~Menzel, S.~Monteil and V.~Niess {\it et al.},
  Phys.\ Rev.\ D {\bf 84} (2011) 033005
  [arXiv:1106.4041 [hep-ph]].


\bibitem{Lenz:2010gu}
  A.~Lenz, U.~Nierste, J.~Charles, S.~Descotes-Genon, A.~Jantsch, C.~Kaufhold, H.~Lacker and S.~Monteil {\it et al.},
  Phys.\ Rev.\ D {\bf 83} (2011) 036004
  [arXiv:1008.1593 [hep-ph]].

\bibitem{newdata}
  I.~Adachi {\it et al.}  [Belle Collaboration],
  arXiv:1208.4678 [hep-ex].
%
  R.~Aaij {\it et al.}  [LHCb Collaboration],
  Phys.\ Rev.\ Lett.\  {\bf 108} (2012) 101803
  [arXiv:1112.3183 [hep-ex]].
%
  R.~Aaij {\it et al.}  [LHCb Collaboration],
  Phys.\ Lett.\ B {\bf 713} (2012) 378
  [arXiv:1204.5675 [hep-ex]].
 %
  R.~Aaij {\it et al.}  [LHCb Collaboration],
  Phys.\ Rev.\ Lett.\  {\bf 108} (2012) 241801
  [arXiv:1202.4717 [hep-ex]].


\bibitem{Aaij:2012ke}
  R.~Aaij {\it et al.}  [LHCb Collaboration],
  arXiv:1211.6093 [hep-ex].

\bibitem{Jung:2012mp}
  M.~Jung,
  Phys.\ Rev.\ D {\bf 86} (2012) 053008
  [arXiv:1206.2050 [hep-ph]].

\bibitem{Bigi:1981qs}
  I.~I.~Y.~Bigi and A.~I.~Sanda,
  Nucl.\ Phys.\ B {\bf 193} (1981) 85.

\bibitem{PenEstimates}  
  H.~Boos, T.~Mannel and J.~Reuter,
  Phys.\ Rev.\ D {\bf 70} (2004) 036006
  [hep-ph/0403085].
  %
  H.~-n.~Li and S.~Mishima,
  JHEP {\bf 0703} (2007) 009
  [hep-ph/0610120].
  %
  M.~Gronau and J.~L.~Rosner,
  Phys.\ Lett.\ B {\bf 672} (2009) 349
  [arXiv:0812.4796 [hep-ph]].


\bibitem{Beneke:2000ry}
  M.~Beneke, G.~Buchalla, M.~Neubert and C.~T.~Sachrajda,
  Nucl.\ Phys.\ B {\bf 591} (2000) 313
  [hep-ph/0006124].

\bibitem{PenUspin}
  R.~Fleischer,
  Eur.\ Phys.\ J.\ C {\bf 10} (1999) 299
  [hep-ph/9903455].
 %
  M.~Ciuchini, M.~Pierini and L.~Silvestrini,
  Phys.\ Rev.\ Lett.\  {\bf 95} (2005) 221804
  [hep-ph/0507290].
 %
  M.~Ciuchini, M.~Pierini and L.~Silvestrini,
  arXiv:1102.0392 [hep-ph].
%
  S.~Faller, M.~Jung, R.~Fleischer and T.~Mannel,
  Phys.\ Rev.\ D {\bf 79} (2009) 014030
  [arXiv:0809.0842 [hep-ph]].
%
  S.~Faller, R.~Fleischer and T.~Mannel,
  Phys.\ Rev.\ D {\bf 79} (2009) 014005
  [arXiv:0810.4248 [hep-ph]].
 
\bibitem{NewDataJPsiP}
  T.~Aaltonen {\it et al.}  [CDF Collaboration],
  Phys.\ Rev.\ D {\bf 83} (2011) 052012
  [arXiv:1102.1961 [hep-ex]].
%
  Aaij {\it et al.}  [LHCb Collaboration],
  Phys.\ Lett.\ B {\bf 713} (2012) 172
  [arXiv:1205.0934 [hep-ex]].
%
  R.~Aaij {\it et al.}  [LHCb Collaboration],
  Phys.\ Rev.\ D {\bf 85} (2012) 091105
  [arXiv:1203.3592 [hep-ex]].
  
\bibitem{SU3breakingmodind}
  M.~J.~Savage,
  Phys.\ Lett.\ B {\bf 257} (1991) 414.
  %
  M.~Gronau, O.~F.~Hernandez, D.~London and J.~L.~Rosner,
  Phys.\ Rev.\ D {\bf 52} (1995) 6356
  [hep-ph/9504326].
 %
  B.~Grinstein and R.~F.~Lebed,
  Phys.\ Rev.\ D {\bf 53} (1996) 6344
  [hep-ph/9602218].
  %
  M.~Jung and T.~Mannel,
  Phys.\ Rev.\ D {\bf 80} (2009) 116002
  [arXiv:0907.0117 [hep-ph]].
%

\bibitem{HFAGPDG}
  Y.~Amhis {\it et al.}  [Heavy Flavor Averaging Group Collaboration],
  arXiv:1207.1158 [hep-ex], and online update at http://www.slac.stanford.edu/xorg/hfag.
%
  J.~Beringer {\it et al.}  (Particle Data Group),
  Phys.\ Rev.\ D {\bf 86} (2012) 010001.

\bibitem{Fleischer:proc}
R.~Fleischer, these proceedings.

\bibitem{Lee:2007wd}
  S.~E.~Lee {\it et al.}  [Belle Collaboration],
  Phys.\ Rev.\ D {\bf 77} (2008) 071101
  [arXiv:0708.0304 [hep-ex]].

\bibitem{Aubert:2008bs}
  B.~Aubert {\it et al.}  [BABAR Collaboration],
  Phys.\ Rev.\ Lett.\  {\bf 101} (2008) 021801
  [arXiv:0804.0896 [hep-ex]].

\bibitem{newdataBtoJPsiV}
  R.~Aaij {\it et al.}  [LHCb Collaboration],
  Phys.\ Rev.\ D {\bf 86} (2012) 071102
  [arXiv:1208.0738 [hep-ex]].
  %
  R.~Aaij {\it et al.}  [LHCb Collaboration],
  arXiv:1210.2631 [hep-ex].
\end{thebibliography}
\end{document}